
\def\figflag{I}                                                           %
\input harvmac
\def\Title#1#2{\rightline{#1}\ifx\answ\bigans\nopagenumbers\pageno0\vskip1in
\else\pageno1\vskip.8in\fi \centerline{\titlefont #2}\vskip .5in}

%
%
\def\Fig#1{Fig.~\the\figno\xdef#1{Fig.~\the\figno}\global\advance\figno
 by1}
\def\figI{I}
%
%
\newdimen\tempszb \newdimen\tempszc \newdimen\tempszd \newdimen\tempsze
\ifx\figflag\figI
\input epsf
%
\def\epsfsize#1#2{\expandafter\epsfxsize{
 \tempszb=#1 \tempszd=#2 \tempsze=\epsfxsize
     \tempszc=\tempszb \divide\tempszc\tempszd
     \tempsze=\epsfysize \multiply\tempsze\tempszc
     \multiply\tempszc\tempszd \advance\tempszb-\tempszc
     \tempszc=\epsfysize
     \loop \advance\tempszb\tempszb \divide\tempszc 2
     \ifnum\tempszc>0
        \ifnum\tempszb<\tempszd\else
           \advance\tempszb-\tempszd \advance\tempsze\tempszc \fi
     \repeat
\ifnum\tempsze>\hsize\global\epsfxsize=\hsize\global\epsfysize=0pt\else\fi}}
\epsfverbosetrue
\fi
%
%
%
%
\def\ifigure#1#2#3#4{
\midinsert
\vbox to #4truein{\ifx\figflag\figI
\vfil\centerline{\epsfysize=#4truein\epsfbox{#3}}\fi}
\narrower\narrower\noindent{\footnotefont
{\bf #1:}  #2\par}
\endinsert
}
%
%
\font\ticp=cmcsc10
\def\calm{{\cal M}}

\def\ajou#1&#2(#3){\ \sl#1\bf#2\rm(19#3)}
\def\frac#1#2{{#1 \over #2}}
\def\grr{g_{rr}}
\def\gtt{g_{tt}}
\def\hf{{1\over2}}

\def\rst{r^*}

\def\p+{{\partial_+}}

\def\delsl{\,\raise.15ex\hbox{/}\mkern-13.5mu \nabla}
\def\Ssl{{\,\raise.15ex\hbox{/}\mkern-10.5mu S}}
\def\lpl{l_{\rm pl}}
\def\tpl{t_{\rm pl}}
\def\mpl{m_{\rm pl}}
\def\calo{{\cal O}}
\def\cald{{\cal D}}
\def\tila{{\tilde A}}
\def\Xdot{{\dot X}}
\def\EF{Eddington-Finkelstein}
\def\caln{{\cal N}}
\def\hc{{\rm h.c.}}
%
%
\lref\deAl{S.P. deAlwis, ``Quantization of a theory of 2d dilaton
gravity,''\ajou Phys. Lett. &B289 (92) 278, hep-th/9205069\semi
``Black hole physics from Liouville theory,''\ajou
Phys. Lett. &B300 (93) 330, hep-th/9206020.}
\lref\StTr{A. Strominger and S. Trivedi, ``Information consumption by
Reissner-Nordstrom Black Holes,'' ITP/Caltech preprint
NSF-ITP-93-15=CALT-68-1851, hep-th/9302080.}
\lref\GaSt{D. Garfinkle and A. Strominger, ``Semiclassical Wheeler wormhole
production,''\ajou Phys. Lett. &B256 (91) 146.}
\lref\HVer{H. Verlinde, ``Black
holes and strings in two dimensions,'' Princeton preprint PUPT-1303,
to appear in the proceedings of the Sixth Marcel Grossman Meeting.}
\lref\MTW{C.W. Misner, K.S. Thorne, and J.A. Wheeler, {\sl Gravitation}
(W.H. Freeman, 1973), pp. 828-832.}
\lref\VeVe{E. Verlinde and H. Verlinde, `` A unitary S-matrix for 2d black
hole formation and evaporation,'' PUPT-1380=IASSNS-HEP-93/8.}
\lref\Wilc{F. Wilczek, ``Quantum purity at a small price: easing a black
hole paradox,'' IASSNS-HEP-93/12, hep-th/9302096.}
\lref\CaWi{R.D. Carlitz and R.S. Willey, ``Reflections on moving
mirrors,''\ajou Phys. Rev. &D36 (87) 2327; ``Lifetime of a black
hole,''\ajou Phys. Rev. &D36 (87) 2336.}
\lref\Schw{J. Schwinger, ``On gauge invariance and vacuum
polarization,''\ajou Phys. Rev. &82 (51) 664.}
\lref\AfMa{I.K. Affleck and N.S. Manton, ``Monopole pair production in a
magnetic field,''\ajou Nucl. Phys. &B194 (82) 38.}
\lref\AAM{I.K. Affleck, O. Alvarez, and N.S. Manton,
``Pair production at strong
coupling in weak external fields,''\ajou Nucl. Phys. &B197 (82) 509.}
\lref\BOS{T. Banks, M. O'Loughlin, and A. Strominger, ``Black hole remnants
and the information puzzle,'' hep-th/9211030, {\sl Phys. Rev. D} to appear.}
\lref\Morg{D. Morgan, ``Black holes in cutoff gravity,"\ajou Phys.
Rev. &D43 (91) 3144.}
\lref\BaOl{T. Banks and M. O'Loughlin, ``Classical and quantum production
of cornucopions at energies below $10^{18}$ GeV,'' Rutgers preprint
RU-92-14.}
\lref\DXBH{S.B. Giddings and A. Strominger, ``Dynamics of Extremal Black
Holes,''\ajou Phys. Rev. &D46 (92) 627, hep-th/9202004.}
\lref\GHS{D. Garfinkle, G. Horowitz, and A. Strominger, ``Charged black holes
in string theory,''\ajou Phys. Rev. &D43 (91) 3140, erratum\ajou Phys. Rev.
& D45 (92) 3888.}
\lref\GiMa{G.W. Gibbons and K. Maeda, ``Black holes and membranes in
higher-dimensional theories with dilaton fields,''\ajou Nucl. Phys. &B298
(88) 741.}
\lref\Pres{J. Preskill, ``Do black holes destroy information?'' Caltech
preprint CALT-68-1819, hep-th/9209058.}
\lref\ACN{Y. Aharonov, A. Casher, and S. Nussinov, ``The unitarity
puzzle and Planck mass stable particles,"\ajou Phys. Lett. &B191 (87)
51.}
\lref\BHMR{S.B. Giddings, ``Black holes and massive remnants,''\ajou Phys.
Rev. &D46 (92) 1347, hep-th/9203059.}
\lref\HaSt{J.A. Harvey and A. Strominger, ``Quantum aspects of black
holes,'' preprint EFI-92-41, hep-th/9209055, to appear in the proceedings
of the 1992 TASI Summer School in Boulder, Colorado.}
\lref\Erice{S.B. Giddings, ``Toy models for black hole evaporation,''
UCSBTH-92-36, hep-th/9209113, to appear in the proceedings of the
International Workshop of Theoretical Physics, 6th Session, June 1992,
Erice, Italy.}
\lref\RSTii{J.G. Russo, L. Susskind, and L. Thorlacius, ``The
Endpoint of Hawking Evaporation,'' Stanford preprint SU-ITP-92-17.}
\lref\HawkTd{S.W. Hawking, ``Evaporation of two dimensional black holes,''
CalTech preprint CALT-68-1774, hep-th@xxx/9203052.}
\lref\CGHS{C.G. Callan, S.B. Giddings, J.A. Harvey, and A. Strominger,
``Evanescent black holes,"\ajou Phys. Rev. &D45 (92) R1005.}
\lref\RST{J.G. Russo, L. Susskind, and L. Thorlacius, ``Black hole
evaporation in 1+1 dimensions,'' Stanford preprint SU-ITP-92-4.}
\lref\BDDO{T. Banks, A. Dabholkar, M.R. Douglas, and M O'Loughlin, ``Are
horned particles the climax of Hawking evaporation?'' \ajou Phys. Rev.
&D45 (92) 3607.}
\lref\BGHS{B. Birnir, S.B. Giddings, J.A. Harvey, and A. Strominger,
``Quantum black holes,'' UCSB/Chicago preprint UCSBTH-92-08=EFI-92-16,
hep-th@xxx/\-9203042.}
\lref\BiCa{A. Bilal and C. Callan, ``Liouville models of black hole
evaporation,'' Princeton preprint PUPT-1320, hep-th/9205089.}
\lref\BDDO{T. Banks, A. Dabholkar, M.R. Douglas, and M O'Loughlin, ``Are
horned particles the climax of Hawking evaporation?'' \ajou Phys. Rev.
&D45 (92) 3607.}
\lref\HawkEvap{S.W. Hawking, ``Particle creation by black
holes,"\ajou Comm. Math. Phys. &43 (75) 199.}
\lref\HawkUnc{S.W. Hawking, ``The unpredictability of quantum
gravity,''\ajou Comm. Math. Phys &87 (82) 395.}
\lref\BPS{T. Banks, M.E. Peskin, and L. Susskind, ``Difficulties for the
evolution of pure states into mixed states,''\ajou Nucl. Phys. &B244 (84)
125.}
\lref\Sred{M. Srednicki, ``Is purity eternal?,'' UCSB preprint
UCSBTH-92-22, hep-th/9206056.}
\lref\SuTh{L. Susskind and L. Thorlacious, ``Hawking radiation and
back-reaction,'' Stanford preprint SU-ITP-92-12, hep-th@xxx/9203054.}

\Title{\vbox{\baselineskip12pt\hbox{UCSBTH-93-08}\hbox{hep-th@xxx/9304027}
}}
{\vbox{\centerline {Constraints on Black Hole Remnants}
}}
\centerline{{\ticp Steven B. Giddings}\footnote{$^\dagger$}
{Email addresses:
giddings@denali.physics.ucsb.edu, steve@voodoo.bitnet.}
}
\vskip.1in
\centerline{\sl Department of Physics}
\centerline{\sl University of California}
\centerline{\sl Santa Barbara, CA 93106-9530}

\bigskip
\centerline{\bf Abstract}
One possible fate of information lost to black holes is its preservation in
black hole remnants.  It is argued that a type of effective field theory
describes such remnants (generically referred to as informons).  The
general structure of such a theory is investigated and the infinite pair
production problem is revisited.  A toy model for remnants clarifies some
of the basic issues; in particular, infinite remnant production is not
suppressed simply by the large internal volumes as proposed in cornucopion
scenarios.  Criteria for avoiding infinite production are stated in terms
of couplings in the effective theory.  Such instabilities remain a
problem barring what would be described in that theory as
a strong coupling conspiracy.  The relation to euclidean calculations of
cornucopion production is sketched, and potential flaws in that analysis
are outlined.  However, it is quite plausible that pair production of
ordinary black holes ({\it e.g.} Reissner N\"ordstrom or others) is
suppressed due to strong effective couplings.  It also remains an open
possibility that a microscopic dynamics can be found yielding an
appropriate strongly coupled effective theory of neutral informons without
infinite pair production.

\Date{}

\newsec{Introduction}

The physics of black holes has spawned a serious clash between the basic
principles of general relativity and quantum mechanics.  The problem is
illustrated by considering the evolution of a system of mass $M$
undergoing
gravitational collapse to a black hole.
In the far past one can take the gravitational interactions to be weak
and the system to be in a pure quantum state.  The system then collapses
to form a horizon, and begins to emit Hawking radiation \refs{\HawkEvap}.
This radiation is approximately thermal.  If we imagine that the black
hole completely evaporates the resulting Hawking radiation is
apparently described by a mixed state.  One then concludes that in the
context of black-hole formation and evaporation, pure states evolve into
mixed states: unitary quantum mechanical evolution fails and an amount
of quantum information $\propto M^2/\mpl^2$ (where $\mpl$ is the Planck
mass) is lost.

Following this logic, Hawking \refs{\HawkUnc} proposed that the laws of
physics are fundamentally nonunitary, and are described by an operator
$\Ssl$ that linearly maps density matrices to density matrices.  However,
this proposal suffers several flaws.  First, non-unitarity is
distasteful and should be contemplated only as a last resort.  Second,
and more concretely, after careful examination of such evolution, Banks,
Peskin, and Susskind \refs{\BPS}, and Srednicki \refs{\Sred} have
argued that failure of unitarity indicates failure of energy
conservation at the same level.  This would have dire consequences.

We are therefore confronted with the puzzle of explaining the fate of
information that falls into a black hole.\foot{For recent reviews of the
information problem see, {\it e.g.}, \refs{\BHMR\Pres\HaSt-\Erice}.}
One obvious possibility is
that it escapes in the evaporation process; in fact, Hawking's
calculation is valid only to the extent that one neglects the
backreaction of the radiation and other quantum effects.  However,
recent investigation of the analogous problem for two-dimensional black
holes \refs{\CGHS\HVer\RST\BDDO\SuTh\HawkTd\BGHS\deAl\BiCa-\RSTii,\HaSt,\Erice}
has allowed greater analytical control.  This analysis seems to indicate
that information can escape in Hawking radiation only as a result of
introduction of drastically new physics; in particular violation of the
equivalence principle at low energy/weak curvature seems to be required.

A different possibility is that the information is emitted once the
black hole reaches the Planck size, where the preceding analysis is no
longer trustworthy. This would, however, require emission of an
unboundedly large amount of information (corresponding to the
possibility of an arbitrarily large initial black hole) in the remaining
energy $E\sim \mpl$. That must take a long time
\refs{\HawkUnc,\ACN,\BHMR,\Pres}, implying a long-lived black-hole remnant.

Such remnants must be of Planck size and have an infinite number of
states to allow them an unbounded information content; since this is
their primary characteristic (and to distinguish them from larger remnants)
they will be generically referred to as
{\it informons}.  They have long been postulated in the abstract
\refs{\HawkUnc, \ACN} as one solution to the information problem, but
suffer a potentially serious flaw: despite the nearly infinitesimal
probability
for producing any given informon in ordinary physical processes, their
infinite degeneracy would seem to imply an infinite inclusive production
rate.

A more concrete realization of such objects in the context of charged black
holes has arisen in recent works;
one informon candidate is the extremal charged
dilatonic black hole of \refs{\GiMa, \GHS}. At the extremal value of the
mass (determined by the charge) the spatial geometry of the solution is
that of an infinitely-long throat attached to an asymptotically flat
region.  A natural conjecture is that the infinite number of states
could correspond to excitations of the infinite throat.  This
possibility was investigated in the two-dimensional reduction of this
theory to excitations moving along the throat in \refs{\CGHS}. The
connection to the full four-dimensional theory was later elaborated in
\refs{\BDDO, \DXBH}; the name {\it cornucopions} was coined for such
objects in \refs{\BDDO}.

In addition to the generic problem of infinite
pair production (to which we will
return momentarily), the specific cornucopion proposal has two other
potentially serious flaws.  First, the basic scenario has suffered
difficulties due to semiclassical
singularities  \refs{\RST, \BDDO}.  One may however take the attitude that
a better understanding of physics in strong coupling could remedy this.
Another problem is that cornucopions seem to require a charge to
stabilize the throat, so oppositely charged cornucopions could
annihilate, resulting in loss of their information.  To avoid this a
rationale for long-lived neutral remnants must be found.

One attractive rationale is the hypothesis that the underlying quantum
theory of gravity places a bound on information content within a given
volume, roughly corresponding to one state per Planck volume.  Although
it is a challenge to find a dynamical implementation of
this proposal, it is a
plausible feature of a quantum theory of gravity and in particular is
hinted at in the properties of string theory.  In \refs{\BHMR} this
assumption was argued to imply either the formation of massive remnants
or of Planck size remnants with large internal volume.\foot{A crude
model of the dynamics of possibly similar objects was studied previously
in \refs{\Morg}.}
The former would seem to require new physics at weak curvatures and runs
afoul of causality.  If the latter can be made physically acceptable it is
thus preferable.

Therefore it is plausible that the problems specific to cornucopions or
other related objects could be resolved with an improved understanding of
gravity near the Planck scale.  That leaves the problem of infinite pair
production of informons.

Recent suggestions in this direction have been made by Banks,
O'loughlin, and Strominger \refs{\BaOl, \BOS}. They argue that the
infinite near-degeneracy of the cornucopions implies that one must be
careful in treating them by low-energy effective field theory, casting
suspicion on calculations of production rates.  In particular \refs{\BaOl}
gives
heuristic arguments for suppression of cornucopion production due to their
large volumes; the infinite number of cornucopion states are very far away
in the internal space and thus cannot be excited in a finite time by
internal causality.
Ref.~\refs{\BOS} furthermore reconsiders Schwinger/Affleck/Manton
\refs{\Schw,\AfMa} pair production in an electromagnetic field.  An
approximate
instanton is found for the cornucopion version of this
process.\foot{This is similar to the instanton for pair production of
Reissner-Nordstrom black holes discussed in \refs{\GaSt}.}  In the
corresponding euclidean geometry a horizon forms as a result of the
interaction energy required to accelerate the throat.  The throat is
thus cut off at finite distance and the euclidean action is finite.
Ref.~\refs{\BOS} advocates that the finite answer results from the
failure of the infinite number of states ``behind the horizon'' to
contribute to the calculation.

There is an objection to these arguments.  If there are an infinite number of
``ground states'' of the cornucopion, then there should be an infinite
number of excited black-hole states that result from adding the
interaction energy. It is not clear why this infinite degeneracy
shouldn't be included in the calculation; instead the calculation
appears to include a factor roughly equal to the Hawking-Bekenstein
entropy of the resulting excited black hole.
To resolve these issues a more thorough analysis is required.
In particular one would like
to relate the calculation in \BOS\ to standard derivations of pair
production.

Such a relation arises through an effective field theory for informons.
Indeed, informons appear as point-like
particles to an external
observer.  Furthermore in general one's
analysis must be carried out in the absence of a detailed
knowledge of Planck scale physics.
It then seems reasonable to adopt the pragmatic principle that {\it anything
plausible can happen at the Planck scale}, and thus to characterize informon
physics only in general terms.
In particular, we do not know that informons
have large volume
interiors.  They must, however, have an infinite spectrum below an
energy of order $\mpl$ to solve the information problem.
Therefore it is useful to introduce an
effective field theory which includes the possibility of
infinite informon degeneracy and in which pair creation and other issues
can be investigated.  A particular case of such a theory is that arising
from cornucopions.

In outline, this paper will first describe such an effective theory and its
coupling to external fields.  Next in section three
a toy model for informons will be constructed, and heuristic arguments
based on the large internal volume in this model will be offered that
infinite production does not occur.  These arguments are false, as the more
careful analysis of section four shows.  This section also states in
specific terms the infinite production problem.  The following section
investigates the possibility that finite lifetimes for remnants could
change the infinite production rate.  Section six makes a comparison with
the finite calculations for pair production of excited cornucopions,
as appeared in
\refs{\BOS}, and a possible source of discrepancy is traced.
Also
discussed is pair production of ordinary black holes:  if black holes do
not lose information and thus have infinitely many internal states, one may
worry that they are infinitely produced.  Conclusions and speculations appear
in section seven.

\newsec{Effective theories for informons}

Suppose we assume that information is not emitted nor new physics
encountered during Hawking
evaporation before the mass and radius of the black hole reach the Planck
scale, and also that information is not destroyed.
Black holes then must leave  remnants with
mass comparable to $\mpl$ and radius of order $\lpl$.
These must have an infinite number of ``internal states''
to allow them to store the information from an arbitrarily large initial
black hole.  These informon states are written
\eqn\remlabel{ \vert k,A\rangle}
where $k$ is the momentum and $A$ labels the states in the Hilbert space
corresponding to a single informon.
Standard field theory arguments based on causality and Lorentz invariance
at distances $\gg \lpl$ then imply that antiremnant states must
also exist, and that informons are described by a field
\eqn\inffield{I_A(x) = \int {d^3 k\over (2\pi)^3 2\omega_k}\left[ I_A(k)
e^{-ik\cdot x} + I_A^\dagger(k)e^{ik\cdot x}\right]\ .}
Free informon propagation is then governed by the action\foot{For
simplicity we take the informons to be scalar, although the labels
$A$ could equally well include spin.  The following discussion would
then generalize.}
\eqn\informact{S_K = \int d^4x \sum_A \left[-\hf \left(\partial
I_A\right)^2 -\hf m_A^2 I_A^2\right]\ .}

We also must describe interactions with other fields.  For simplicity first
consider
couplings to an external real scalar field $\phi(x)$.
The general three-point action is
\eqn\threept{S_\phi = -\int d^4x d^4y d^4z \sum_{AB} I_A(x) \langle A|
{\tilde \calo}(x,y;z) |B\rangle I_B(y) \phi(z)}
where ${\tilde \calo}(x,y;z)$ is an operator acting on the
informon Hilbert space as
well as on the coordinates $x,y,z$.  The Planck size of the informon
implies that the interaction is local on larger scales, so
\eqn\newop{{\tilde \calo}(x,y;z) =\calo(x)\delta(x-y)\delta(x-z)}
up to terms falling exponentially at long distances.  This gives
\eqn\localint{S_\phi = -\int d^4x \phi(x) \sum_{AB} I_A(x)\langle A|
\calo(x)|B\rangle I_B(x)\ .}
or, in momentum space,
\eqn\momint{-\int d^4k d^4q \phi(q)\sum_{AB} I_A(k+q) I_B(k) \langle A|
\calo(k^2, q^2, k\cdot q)|B\rangle \ .}
Eq.~\momint\ may
be extended to higher-order interactions.  In general the informon
interacts with its surroundings through a sum of such terms.

The generalizations for couplings to gravity or electromagnetism are
straightforward.  For a weak electromagnetic field the interaction is
likewise of
the form
\eqn\emcoup{S_{\rm em}
= -\int d^4x  A_\mu \sum_{AB}\left[ I_A^*(x) \langle A|{\cal J}^\mu
|B\rangle I_B(x)+ \hc\right]}
where ${\cal J}^\mu$ is an operator that must reduce to the standard
electromagnetic current at zero momentum transfer.  This may be
supplemented by higher order couplings, both as required by gauge
invariance and otherwise.
The general such action for informons of charge $q$
coupled to electromagnetism is then
\eqn\infact{\eqalign{ S[I]=\int d^4x\Biggl\{&  \sum_A
\left[-\left|(\partial_\mu +iqA_\mu)I_A\right|^2- m_A^2
 |I_A|^2\right] \cr &-
\left[\sum_i{\widehat\calo}_i[A_\mu] \sum_{AB} I_A^*(x)
\langle {\bar A}| \calo_i(x^\mu, -i\partial_\nu)
|B\rangle I_B(x)  +{\rm h.c.}\right]\Biggr\} }}
where ${\widehat\calo}_i[A_\mu]$, $\calo_i$  are local operators
acting on the electromagnetic field or internal Hilbert space,
respectively.  ($\calo_i$ may also include derivatives acting on
informon fields, as indicated.)
Likewise for a weak gravitational field, $g_{\mu\nu}=
\eta_{\mu\nu} + h_{\mu\nu}$, the coupling is of the form
\eqn\gravcoup{S_g = -\int d^4 x h_{\mu\nu} \sum_{AB} I_A(x)
\langle A|{\cal T}^{\mu\nu}
|B\rangle I_B(x)}
where again ${\cal T}^{\mu\nu}$ should reduce to the standard stress tensor
at zero momentum transfer; with inclusion of higher-order
interactions this becomes
\eqn\gravhc{\eqalign{
S_g[I]= \int d^4x& \sqrt{-g}\Biggl[ \sum_A {1\over2}\left(- \nabla_\mu I_A
\nabla^\mu I_A- m_A^2 I_A^2 \right)\cr&- \sum_i
{\widehat \calo}_i [g_{\mu\nu}]\sum_{AB}
I_A(x) \langle A| \calo_i(x^\mu, -i\partial_\nu) |B\rangle I_B(x)
\Biggr]\ .}}

It is worth emphasizing the extent to which such theories are
effective field theories in the conventional sense.  First, since
informons
resulting from evaporation of a neutral black hole should have size
$\sim\lpl$ they have excitations with planckian frequencies and
wavenumbers.  These modes are integrated out in writing the
effective field theory.  However, informons
also have an infinite number of low frequency internal modes.
These states are not integrated out but instead are explicitly
accounted for.

One might object that at least for cornucopions or other similar scenarios
where the large
number of modes arise due to a large internal volume it is
awkward to describe the internal states in this fashion.   In that
case one has a clear interpretation of the states in terms of modes
propagating on the throat, and no other description seems necessary.
However, there are two separate reasons to describe the internal
states in this general fashion.  First, for cornucopions infalling
states eventually form black holes, and then the interpretation of
the resulting system in terms of quantum states becomes fogged.
Although the precise description of the quantum
states is not known, it is nonetheless
important to make explicit the assumption that
there is some spectrum of quantum states being excited.
This is in contrast to information loss.  Secondly,
and more importantly,
although it is true that there is a semiclassical description
(of limited validity) and thus a partial picture of the
states for large cornucopions, we can hardly expect this to be true
for Planck sized informons.
Even if they have large volume interiors connected
through Planck scale throats this is impossible for a long distance
observer to explore directly without squeezing through the Planck
scale throat.  More generally one may expect a rather complicated
planckian dynamics that does not correspond to large volume but may
nonetheless (or may not)
have some similar features encoded in the spectrum
and interactions.
In any case, we make the fundamental assumption that
whatever the dynamics of quantum gravity is, it falls into the
framework of unitary evolution of quantum states so that information
is preserved.

In the absence of an understanding of quantum gravity we must therefore
adopt rather modest goals.  These are to find whether there is any
possible spectrum of informon states and any possible set of
interactions of these states with external fields that allows a
solution of the black hole information problem.  Without
understanding the microscopic dynamics, the best one can do is
formulate this question within the framework of an effective theory
for informons such has been described.

\newsec{A toy model and a false argument}

One can readily construct toy models for the internal Hilbert space of an
informon that serve as examples of the above general framework.
In particular, the perceived benefits of having an infinite
volume informon may be implemented by choosing a spectrum and set of
interactions of the appropriate form.  It is irrelevant to the outside
observer whether the internal space is ``real.''

A
particularly simple model arises by taking the internal Hilbert
space to be the Fock space of a single scalar field $f$ propagating on the
semi-infinite line $(0,\infty)$.  The action is
\eqn\Intact{S_f = -\hf \int d^2\sigma \left( \nabla f\right)^2 }
where $\sigma,\tau$ are internal
spacetime coordinates,
and  some boundary conditions, {\it e.g.} Neumann, are specified at
$\sigma=0$.
The states of the informon are
labeled by the occupation numbers of the $f$ eigenmodes,
\eqn\onetwo{I_A \leftrightarrow| A\rangle = |\{n_\kappa\}\rangle =
\prod\limits_k {\left(f^\dagger_\kappa\right)^{n_\kappa}\over
\sqrt{n_\kappa!}} |0\rangle}
where
\eqn\fexp{f(\tau,\sigma) = \int_0^\infty
{d\kappa\over 4\pi\kappa} \left(f_\kappa
e^{-i\kappa \tau} + f^\dagger_\kappa
e^{i\kappa \tau}\right) \cos \kappa\sigma\ .}
It is convenient to take the mass of the informon to be $m_A^2 = m_0^2 +
2H_A$, where $m_0$ is a mass of order $\mpl$ and
$H_A$ is the hamiltonian of the internal theory,
\eqn\IntHam{H_A = \int_0^\infty d\kappa \kappa n_\kappa\ .}
This supplies a spectrum with an infinite number of states below any mass
$>m_0$.  In order to avoid arbitrarily massive informons we may also remove
from the theory all states with mass
larger than $m_{\rm max}>m_0$.

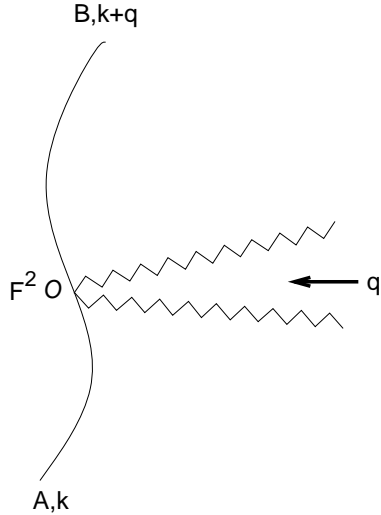
\ifigure{\Fig\intera}{An incoming informon in state $A$ and
with momentum $k$ interacts
with the electromagnetic field through the operator $F^2 \calo$ to produce
an informon in state $B$ with momentum $k+q$.}{cbhr.fig1}{2.75}

Next consider interactions with an external  field; for concreteness take
it to be electromagnetic although other fields could equally well be
considered.
Suppose that the informon is charged and
interacts with the field through the standard minimal coupling terms.
These do not affect the internal structure of the remnant:  they are
diagonal in the informon Hilbert space.  Suppose that there are also
interactions with the field through other operators, e.g.
\eqn\fsq{F^2(q) = \int d^4x e^{-iq\cdot x} F_{\mu\nu}^2(x)\ ,}
and that these are non-diagonal on informon states.
They may be incorporated by temporarily
adopting a first-quantized description for the informon.
The propagator for a free informon of mass $m$ is given by
\eqn\propa{G(x,x')= \int_0^\infty dT \int_x^{x'}
\cald X \exp\left\{i\int_0^T d\tau ({\dot
X}^2 -m^2)\right\}}
where $\tau$ is interpreted as the proper time and ${\dot X} = dX/d\tau$.
We may therefore introduce a coupling to the internal theory, corresponding
to the diagram of fig.~1:
%
%
\eqn\fsqcoup{\eqalign{\int_0^\infty& dT \cald X
e^{i\int_0^{T}d\tau ({\dot X}^2 -m^2)}
e^{ik\cdot X(0)-i(k+q)\cdot X(T)} \int_0^T d\tau'
F_{\mu\nu}^2(X(\tau')) \calo(\tau',0) \cr
&=\int d^4q F^2(q)\int_0^\infty dT \int_0^T d\tau'
e^{i\tau'[k^2+m^2]+i(T-\tau')[(k+q)^2 +m^2]}
\calo(\tau',0) }}
where $\calo(\tau',0)$ is a local operator acting at
$(\tau,\sigma)=(\tau',0)$ in the
internal $f$-theory.  (Free propagators must be truncated from \fsqcoup\ to
define the vertex.)  Such a theory is similar to the effective theories of
combined dimensions introduced to describe cornucopions in
\refs{\DXBH,\BDDO}.

This model shares several features with more fundamental theories of
informons with large internal volumes.  First, there is an infinite
spectrum of low-energy states of the informon arising from the infinite
volume.  Second, any attempt to use electric or magnetic fields to move one
of these objects will inevitably excite the internal state of the informon
through the interaction \fsqcoup.  However, one feature it does not share
with such scenarios is a dynamical geometry.  One
expects that in models with dynamical geometry the dynamics is nonetheless
encoded in
the informon spectrum and interactions, and thus would also be described by
effective theories as in the preceding section.  Although the model of this
section does not capture all aspects of such a theory it is useful for
illustrating relations between infinite volume and pair production.

Within the context of this model arguments similar to
those for finite pair
production of cornucopions\refs{\BDDO,\BaOl}
are easily made.  First, in reference to
thermodynamics problems with informon spectra, it appears that the
infinite ``internal volume'' and the ``causality'' of the internal
theory
imply that equilibration of the thermal ensemble takes a long time.
Furthermore, as emphasized in \refs{\BaOl}, one would expect that the
action to create
configurations that differ from the vacuum over a large (internal)
volume $V$ should
be proportional to $V$.  This reasoning suggests that it should therefore
be difficult to create informon states with excitations arbitrarily far
down the internal line.
Finally and more concretely we can imagine Schwinger pair production of
these in a background electric field.
The instanton arising in the Schwinger process has finite temporal extent;
the
characteristic scale governing
the process is the electric length, $\sim  m_A/qE$.
By
``causality''  of the internal theory it should not possible for states
arbitrarily far down the internal line to be excited by the operator
$\calo$ if it only acts for this finite time
at $\sigma=0$.  Since below $m_{\rm
max}$ there
are only finitely many states that are localized in a finite interval in
$\sigma$, only finitely many remnant states are produced and the total
pair
production rate is finite.  The problem of infinite pair production is
solved.

Unfortunately, the latter arguments are false.

\newsec{Infinite pair production}

The flaw in the above chain of reasoning can be seen for example
by revisiting Schwinger's
calculation as described in \refs{\AAM}.  The decay rate $\Gamma$
of an electric
field into informons
is given by the imaginary part of the euclidean
vacuum-to-vacuum amplitude,
\eqn\decay{V_4 \Gamma = 2 {\rm Im\ ln} \int \cald {\tilde A}_\mu
 \cald I e^{-S[A_\mu]-S[I_A]}}
where $V_4$ is the four-volume, $S[A_\mu]$ is the standard gauge action,
$S[I]$ is the informon action \infact, and the gauge field is
divided into external and fluctuation pieces, $A_\mu = A^0_\mu + \tila_\mu$.
Integrating over $I_A$ gives
\eqn\decaya{V_4 \Gamma = 2 {\rm Im\ ln}
\int \cald {\tilde A}_\mu e^{-S[A_\mu]
-S_{\rm eff}[A_\mu]}}
where
\eqn\effact{ S_{\rm eff}[A_\mu]= \hf {\rm ln\ det} \left\{\left[
 -\left(\partial_\mu + iqA_\mu \right)^2 + m_A^2 \right]\delta_{AB} +
\calm_{AB}\right\}}
and one defines
\eqn\matdef{ \calm_{AB} = \sum_i{\widehat\calo}_i[A_\mu]
\langle {\bar A}|\calo_i|B\rangle+\hc\ .}
Suppose that electromagnetism is weakly coupled (more on this later); then
one may work to leading order in the coupling by simply dropping
$\tila_\mu$.  Using $ln\, det = Tr\, ln$ and the standard Schwinger
proper time trick, the decay rate becomes
\eqn\decayb{ V_4 \Gamma =  {\rm Im} \int_0^\infty {dT\over T} {\rm Tr}\,
e^{-HT}}
%
with ``hamiltonian''
\eqn\ham{\langle {\bar A}|H|B\rangle= \hf\left[
 -\left(\partial_\mu + iqA^0_\mu \right)^2 + m_A^2 \right]\delta_{AB} +\hf
\calm_{AB}(x^\mu,-i\partial_\nu)
\ .}
The trace over positions in \decayb\ can then be rewritten as a functional
integral; an integral over momentum must also appear to accommodate the
derivative dependence in the $\calo_i$.  This yields
\eqn\decayc{ \eqalign{ V_4 \Gamma& = 2{\rm Im} \int_0^\infty {dT\over T} \int
\cald X\cald P
\exp\left\{-\int_0^Td\tau \left[\hf (P+qA^0)^2 -iP_\mu {\dot X}^\mu
\right]\right\}\cr
&\sum_A
\langle{\bar A}| \exp\left\{-{T\over 2} m_A^2 -\hf\int_0^T d\tau
\calm_{AB}(X^\mu,P_\nu)
\right\} |A\rangle\
.}}
For finite production this expression must be finite.

Now the problem of infinite pair production is manifest.  Suppose first
that there is no coupling via operators $\calo_i$.  Suppose also that there
are an infinite number of internal states $|A\rangle$ with masses
$m_A<m_{\rm max}$, for some $m_{\rm max}\sim \mpl$.  Then
\eqn\infsum{\sum_A e^{-T m_A^2} > e^{-m_{\rm max}^2 T} \sum_A 1 =
\infty\,  e^{-m_{\rm max}^2
T}\ .}
This is then inserted into \decayc.  The $P$ integral can be explicitly
performed, and the $T$ integral and $X$ functional integral are given by
the saddlepoint described by the Schwinger instanton.  Recall that this
corresponds to circular motion of the charged particle in the euclidean
continuation of the electric field.  The total
production rate is infinite from
the overall infinite factor.

The infinite production for non-trivial $\calo_i$ can also be seen in
\decayc.  First consider the case of weakly-coupled effective field theory,
in which the couplings through the $\calo_i$ to all states are small.  In
this case the integrals can be evaluated by saddlepoint techniques.  The
$P$ saddlepoint is given by
\eqn\psadd{P_\mu\simeq i{\dot X}_\mu - q A^0_\mu + \cdots}
where neglected terms arise from the $P$ dependence of the $\calo_i$.  The
resulting integral is of the form
\eqn\Xint{ \eqalign{ V_4 \Gamma& = 2{\rm Im} \int_0^\infty {dT\over T} \int
\cald X
e^{-\int_0^Td\tau \left[\hf {\dot X}^2 +iq A^0_\mu {\dot X}^\mu
\right]}
\cr&\sum_A
\langle{\bar A}| \exp\Biggl\{-{T\over2} m_A^2 -\hf\int_0^T d\tau
\calm_{AB}(X^\mu,i{\dot X}_\nu - q A^0_\nu +\cdots)
\Biggr\} |A\rangle\
.}}
Under the assumption of weak coupling, the integrals over
$T$ and $X$ may be approximated by the Schwinger saddlepoint for each $A$.
Near these saddlepoints the
contributions of the $\calo_i$ terms are small.  Therefore
they cannot render the infinite sum finite, and an infinite production rate
follows.

This result could be altered in strong coupling if the
couplings grew arbitrarily large as the states are
enumerated.  However, that would violate tree-level
unitarity
bounds.\foot{Strictly speaking this requires continuation of couplings
from small euclidean frequency to small lorentzian frequency.} This
amounts to saying that the loop effects reduce the effective
couplings to $\calo(1)$.  Therefore the only remaining possibility is that
the couplings are of order one but vary with the internal state of the
informon such that the terms in the sum \decayc\ oscillate and add to a
finite total amplitude.  This would
clearly take precise adjustment of couplings for a given spectrum.

The above result can be explicitly illustrated in
the toy model of the preceding
section.  There the decay rate takes the form
\eqn\decayt{\eqalign{V_4 \Gamma = 2{\rm Im} \int_0^\infty {dT\over T} &
\int \cald X
e^{-\int_0^T d\tau \left[\hf {\dot X}^2 + iqA^0_\mu {\Xdot}^\mu\right]}\cr
&\sum_A
\langle {\bar A}| \exp\left\{-{T\over2}
m_A^2 -\int_0^T d\tau F^2(X(\tau)) \calo(\tau,0)
\right\} |A\rangle\cr
= {\rm Im} \int_0^\infty {dT\over T} &
\int \cald X
e^{-\int_0^T d\tau
\left[\hf {\dot X}^2 + iqA^0_\mu {\Xdot}^\mu\right]}e^{-m_0^2 {T\over 2}}\cr
& \int \cald f e^{-S_f^E
- \int_0^Td\tau F^2(X(\tau)) \calo(\tau,0) }
}}
where the trace over internal states has been rewritten as a functional
integral over the internal $f$-theory, and where $S_f^E$ is the euclidean
version of the free action \Intact.  Now we see directly
the flaw in the previous reasoning.
First off, if the coupling to $F^2$ were turned
off, production would be suppressed by $e^{-H_A T}$, where $H_A$ is the
hamiltonian \IntHam, or equivalently the mass shift.  However, there are
infinitely many states below any mass, resulting in an infinite answer.
Now, by locality, if we turn the
coupling to $F^2$ back on, it can hardly change this infinite answer since
the vast majority of states are far down the line, outside the influence of
the operator $\calo(\tau,0)$.  Production is not suppressed by the large
volume.

These problems likewise extend (pursuing the connection discussed in
Schwinger's classic paper) to the problem of infinite vacuum polarization
and therefore to an infinite and thus nonsensical lagrangian for
electromagnetism.

The same arguments should apply if one considers couplings of informons to
gravity or other fields.  Pair production is only suppressed by the mass
corresponding to the states, so if there are infinitely many states below a
given mass, infinite rates should result.  Couplings to other operators
that affect the internal state of the informon should not alter this unless
these couplings become large for infinitely many of the
internal states and are carefully arranged to oscillate so that they
give a finite production rate.
This greatly restricts the possibility that remnants give
a viable solution to the information problem.

\newsec{Decaying informons}

The above calculation strictly speaking applies only to absolutely stable
informons.  It is also conceivable that remnants decay,
and that this could alter the results.

First let us estimate the decay rate of an informon arising from the
evaporation of a black hole of initial mass $M$.  Such an informon must
have roughly $\caln\sim (M/\mpl)^2$ states excited to encode the the
information.  The characteristic
 energy spacing between these states must therefore be
$\Delta E \sim \mpl/\caln$.  The decay time for one of these states is
$\roughly> 1/\Delta E$.
This may be shown explicitly by considering for example
an s-wave coupling giving decay to massless $\phi$ quanta,
\eqn\scoup{\mpl \int d^4x I_A(x)I_B(x) \phi(x)\ ,}
or equivalently  follows from the observation that it must take a time
$\Delta t \sim 1/\Delta E$ to emit a quantum of energy $\Delta E$.
The characteristic decay time between remnant states is therefore
$\sim \tpl \caln$ and typical decay widths are
$\Gamma_A\sim \mpl/\caln$.
(For the informon to decay completely it must emit
$\caln$ such quanta, taking a time $ \sim \tpl (M/\mpl)^4$,
in accord with the estimate of \refs{\CaWi,\Pres}.)

The resulting mixing
between informon states can effectively be represented by
including a non-trivial mass matrix,
\eqn\massshift{m_A \delta_{AB} \rightarrow m_A \delta_{AB} +\hf \Delta
m_{AB}}
which need not be hermitian.  The elements of $\Delta m_{AB}$ are then of
order $\mpl/\caln$
and make contributions of order $\mpl^2/\caln$ to $m_{AB}^2$.
Reexamining the pair-production rate, \decayc,
we see that a rough criterion for when the finite lifetime becomes
important is if the euclidean time over
which the instanton process takes place satisfies $T\mpl^2/\caln
\roughly>1$.
Therefore for arbitrarily large $\caln$ informon decay
does not make a significant contribution -- the informons decay too slowly.

\newsec{Cornucopion pair production}

Although the above discussion sheds some light on the arguments for finite
production made in \refs{\BDDO,\BaOl} it is yet to be compared with
\refs{\BOS} which claims
to calculate a finite pair production rate for cornucopions.  There it is
pointed out that it is not possible to pair produce cornucopions in their
ground states since any attempt to move them results in excitation of the
cornucopion above extremality and thus formation of a horizon.
Therefore the problem is that of pair producing the resulting black holes.
Ref.~\BOS\ found an approximate euclidean instanton geometry
that describes this
pair production process.   The instanton action is finite, and the
instanton appeared not to include the infinite number of states behind the
horizon that would lead to an infinite degeneracy factor and rate.
However, this argument is quite similar to euclidean arguments that yield
the Bekenstein-Hawking entropy of a black hole; such reasoning
suggests that in fact black holes have only finitely many internal states
in conflict with the presumption that they may contain infinite information.

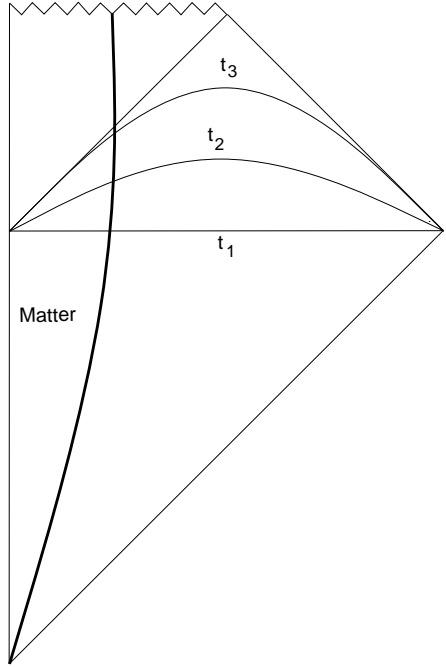
\ifigure{\Fig\pend}{Shown is the time slicing for a
collapsing black hole corresponding
to external Schwarzschild time. As seen by the external observer, the
infalling states never cross the horizon, but are asymptotically frozen at
the horizon.}{cbhr.fig2}{3.5}

In fact the calculation of
\BOS\ apparently neglects the source of the
infinite number of
information-bearing states, and thus sidesteps the issue.
To see this, momentarily consider the simpler problem of forming
a Schwarzschild
black hole by throwing in a large
number of quantum particles.  The Penrose diagram for this process is shown
in fig.~2.  (Temporarily ignore Hawking radiation.)
As described by an observer who uses the standard Schwarzschild
time slicing, the matter never crosses the horizon and the evolution is
unitary.  In this observer's description the information in the
infalling matter is contained in states near the horizon. We
may compare this to a different viewpoint, namely that of ingoing
Eddington-Finkelstein time,\foot{For a concise review of
Eddington-Finkelstein coordinates see \refs{\MTW}.} as shown in fig.~3.
An observer using the \EF\ time slicing would say that the information
crosses the horizon and hits the strong-curvature region.  In either case
the observers count the same number of states (excluding the possibility
that information is truly annihilated at the singularity).  Similar
statements can also be made if Hawking evaporation is allowed, with states
piling up on the effective horizon;\foot{The effective horizon is first
outgoing null ray that hits strong coupling before reaching ${\cal I}^+$.}
since a black hole can begin with an arbitrarily large mass and evaporate
down, the number of such states is infinite.
The moral is
that if we choose to describe processes in terms of external Schwarzschild
time, the ``internal'' states are never really internal: instead they are
frozen at the horizon.

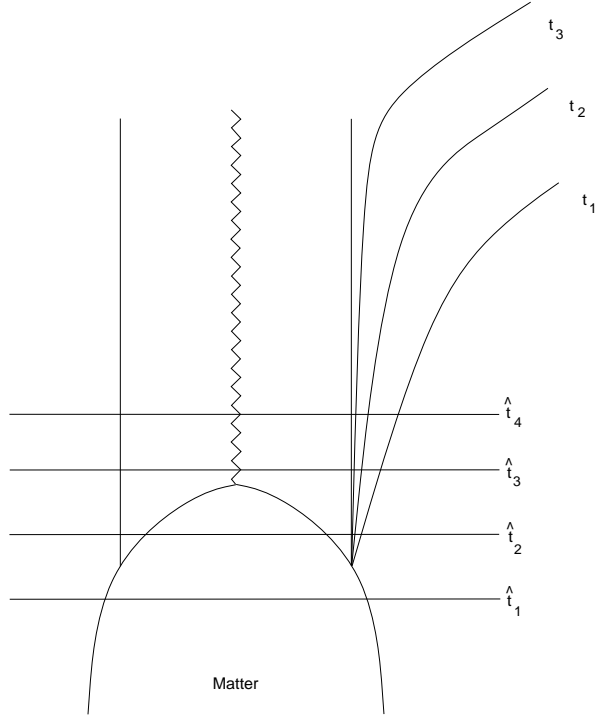
\ifigure{\Fig\eddfink}{The geometry of fig.~2 may equally well be
described using an \EF\
time slicing.  In this time slicing infalling states that the external
observer saw frozen at the horizon instead cross the horizon in finite time
to become internal states of the black hole.}{cbhr.fig3}{3.75}

This behavior can be explicitly illustrated.  Consider for example
a massless scalar field $\psi$
in a static spherically symmetric black-hole geometry,
\eqn\schw{ds^2 = g_{tt} dt^2 + g_{rr}dr^2 + R^2(r) d\Omega_2^2\ .}
For appropriate $g_{tt}$, $g_{rr}$, and $R$ (which we take to asymptote to
flat form at infinity) this specializes to
Schwarzschild or to the near-extremal dilatonic black holes, {\it i.e.}
excited cornucopions.  Mode propagation is most easily discussed in
tortoise coordinates,
\eqn\tort{{dr^*\over dr }= \sqrt{\grr\over -\gtt}\ .}
These range from the horizon at $\rst=-\infty$ to spatial infinity at
$\rst=\infty$.
Expanding in spherical harmonics,
\eqn\spherd{\psi = {u(r,t)\over R} Y_{lm}(\theta,\phi)\ ,}
the action becomes
\eqn\tact{-\int d^4 x \left(\nabla \psi\right)^2 \propto \int dr^* dt
\left[(\partial_t u)^2 - (\partial_{r^*} u)^2 - V(r^*) u^2 \right] }
with effective potential
\eqn\effpot{V(r^*) = -\gtt {l(l+1)\over R^2} +
{\partial_{r^*}^2 R\over R}\
.}
This effective potential vanishes at infinity and at the horizon.

An
incoming mode of a given frequency $\omega$
at infinity takes the form $e^{-i\omega(t+\rst)}$ near the horizon, plus
the part reflected form the barrier.  A wavepacket formed
from these modes will reach $\rst=-\infty$ only at $t=\infty$; as described
in $r$ coordinates it gets asymptotically close to the horizon at infinite
time.   There are an infinite number of such states frozen at the horizon,
arising from the infinite volume in $r^*$, or equivalently from the
infinite number of past histories a black hole of mass $M$ may have had.

These observations may now be applied to pair production of
black holes arising from excitations of cornucopions.
Whereas one could certainly attempt to find a description of black hole
pair production in \EF\ coordinates, it is much simpler to describe the
process in Schwarzschild coordinates.  As emphasized above, these
coordinates, and their euclidean continuation, fail to cover the interior
of the black hole.  However, as described in these coordinates there should
be an infinite number of states at the horizon, corresponding to the
infinite number of possible histories of the black hole.
These would at least na\"\i vely
appear to give an infinite production rate through the
fluctuation determinant about the instanton.

It would be useful to actually compute the fluctuation determinant to
verify this.  In particular one might think
a euclidean momentum cutoff would remove infinitely
many of these states
since they have large wavenumbers near the horizon.  However,
it is not clear that this is sensible:  these modes arise from propagation
of perfectly well-behaved physical states into the black hole.  Similar
outgoing modes are also crucial to the Hawking radiation.  Therefore in
general they should not be truncated from the theory.
If their contribution proves to be negligible that should have a physical
explanation.  Since these are the
information-bearing modes
it would be quite interesting to understand the physics modifying their
contribution.\foot{It is conceivable that a class of cutoffs similar to a
cutoff in Kruskal momentum renders the answer finite and has a physical
interpretation.}

{}From the point of
view of the effective field theory the only possible such rationale is
carefully arranged strong couplings.
However,    as pointed out in \DXBH\
the only low-energy modes propagating on the throat of the cornucopion
are fermion modes.
Their contributions are not expected to make the Schwinger production rate
finite since charged fermions are reflected from the throat.  Rather the
arguments of \BaOl\ rely on propagating modes that excite
the gauge field corresponding to rotations of the throat.
These should then be massive.\foot{The
former point was made by L. Thorlacius, the latter by Banks and
Strominger.}  Their propagation down the throat is exponentially attenuated
and thus they cannot produce a strongly coupled effective field theory.
One might then ask if this is true why do they nonetheless produce a
horizon as described in \BOS.  This presumably occurs due to the unbounded
growth of the coupling constant along the throat; in fact for fixed
asymptotic coupling the horizon forms at arbitrarily large coupling for
arbitrarily weak external field.

The upshot is the formation of the horizon and strong coupling of the
effective field theory both rely not only on {\it strong} ($\calo(1)$)
coupling in the underlying theory of the throat but on {\it unbounded}
coupling.  This makes the argument of \BOS\ for finite production
rather implausible.
In particular if
a strong coupling modification of cornucopions did not have
unbounded couplings but nonetheless had
infinitely many states then the effective field theory should be
weakly coupled
and infinite production results.

On the other hand similar logic can be applied to investigate a
potentially more serious
concern.
If it is true that information is retained in a black hole as it
evaporates, then a black hole of mass (say) 1gm should have an infinite
number of internal states.  Therefore when we describe these in an
effective field theory at energies $E\ll1/1gm$, we might expect to encounter
infinite pair production!

For real black holes, however, there are possible outs.  First, the
modes with energies and angular momenta satisfying $E M \roughly> l$ have
amplitude of order one to penetrate the black hole and thus are strongly
coupled as viewed in the effective theory, in contrast to cornucopions.
Thus what would be described from the effective theory as a strong coupling
conspiracy is quite plausible.  If we considered for example pair
production of extremal Reissner-Nordstrom black holes\GaSt\ this could also
be directly connected to the apparent finiteness of the fluctuation
determinant with an appropriate momentum cutoff.
Furthermore, the
calculation is modified by the Hawking decay of the black
hole, following arguments of the previous section.
The characteristic time for a black hole state of mass $M$ to
decay to a lower state is
given by $\tpl M/\mpl$, in accord with the fact
that the radiation has
characteristic energy $\mpl^2/M$.
Thus from the discussion of the preceding section $\Delta m^2 \sim \mpl^2$
and the rough criterion for the decay to be important is
$T \mpl^2\roughly>1$ where $T$ is the proper time for the appropriate
instanton.  This criterion will be satisfied for an instanton producing a
black hole larger than the Planck mass.  Therefore strong effective
couplings and/or the decay can
substantially suppress the pair production rate.
These arguments then offer a plausible explanation for infinite
production of cornucopions yet finite production of ordinary black
holes.\foot{They also lend credence to the possibility that
Reissner-Nordstrom black holes are viable charged remnants\refs{\StTr}.}

It is however
worth emphasizing the possibility that  strong
couplings and/or short decay times
don't prevent infinite production.  This is a concern since
the rate may be arbitrarily small
for a given black hole state yet infinite due to
the infinite number of possible black hole states.  That raises the
disturbing possibility of infinite production of black holes in
slowly varying macroscopic
fields.   If this were the case  it would be
difficult to see another convincing way out of the catastrophe.  The
possibilities are:

\item{1.}  Although information is not lost by black holes, and black holes
can be assigned internal states, black holes cannot be treated by the type
of effective
field theories being discussed.  However, it is difficult to see how this
could be so, from the reasoning in section 2.
To reiterate:
black holes are localized,
and if they are allowed to have internal states and momenta the only known
description of them that is Lorentz invariant and causal
at long distances is in terms
of effective fields.  One would need a loophole in these arguments.

\item{2.}  Information does not pass the horizon and is reradiated in some
form.  This would likely require new physics and in particular a breakdown
of the equivalence principle at weak curvatures.

\item{3.}  Information is in fact annihilated.  However, this would be
expected to happen only at the singularity.  This phenomenon could
therefore presumably not be discovered by an external observer until the
black hole reached the Planck size.  Such an external observer would thus
still attribute an infinite number of states to the black hole, and
therefore would still apparently
confront infinite pair production of black holes.
One escape would
be annihilation of information at the horizon, but this would again require
a major breakdown of known physics at arbitrarily weak curvatures.

\newsec{Conclusions and Speculations}

If black holes leave remnants with internal states of characteristic size
and mass
$r_{\rm Remnant}$, $m_{\rm Remnant}$,
then these remnants
should be describable in terms of an effective field theory on scales
$E<m_{\rm Remnant},1/r_{\rm Remnant}$.
This follows from the basic principles of quantum
mechanics, macroscopic Lorentz invariance, and causality.
There should be one field
$I_A$ for each allowed remnant internal state $|A\rangle$.  Couplings
of slowly-varying external fields to such informons
are also easily incorporated in such an
effective field theory using the basic technology of form factors.
If the spectrum of informons is infinite and if this effective theory is
weakly coupled then informons will be infinitely produced in processes like
that of Schwinger.  Thus there is
no viable weakly coupled informon (hence
cornucopion) scenario.  It may be possible that this conclusion is altered
for a strongly coupled effective theory of informons, but only
if the couplings are carefully
fixed through what would be described in that theory as
a strong coupling conspiracy.  In particular it is far from
clear that the cornucopion scenario consistently arranges such a conspiracy;
the
finite rates computed in \refs{\BOS} neglected the probable origin of the
infinite number of states and rely on unbounded coupling in the underlying
field theory.
The same logic does not however apply to ordinary black holes
since they are strongly coupled in the effective sense
and also have relatively short decay times.

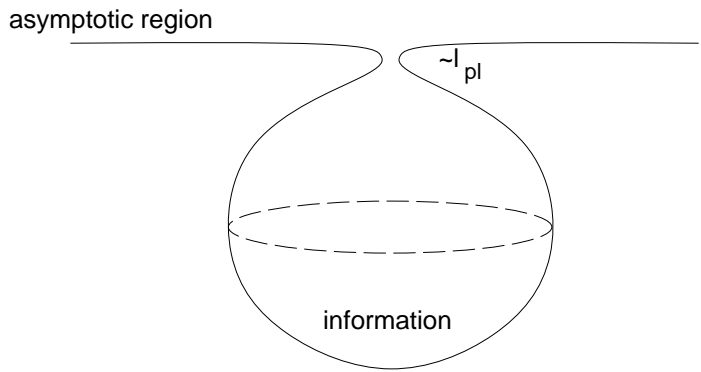
\ifigure{\Fig\infobag}{A (speculative) picture of an informon with a large
internal volume in all dimensions.  Such an
object would have light ``internal'' modes for all external fields, and
possibly could be arranged to have appropriate couplings at the throat to
prevent infinite pair production.}{cbhr.fig4}{2.}

Where does this leave us with the information problem?  Strictly speaking
informons are not yet ruled out since strong coupled effective theories
leave us a
little room for maneuvering.  Given difficulties with other proposed
resolutions of the information conundrum it is certainly worth
investigating the possibility of viable scenarios.  It would seem that at a
minimum all fields that can pair produce informons should be effectively
strongly
coupled to avoid infinite production.  This requires
that all such fields couple strongly to light modes of the informon.  One
way to imagine such a possibility is if informons consist of internal
regions that are not just large in one direction but in all directions.  In
this case there would indeed be light modes internal to the informon.  The
large volume could be connected through a Planck-sized throat, and with
appropriate dynamics the couplings might be arranged to be sufficiently
strong.  Perhaps such a picture (see fig.~4) would yield a finite
production rate.  The challenge is to find a viable dynamics
that describes such an object, or other realizations of informon scenarios
that evade infinite production.  If quantum gravity indeed includes such
states with Planck masses and infinite spectra this may be
a deep clue towards its structure.

\bigskip\bigskip\centerline{{\bf Acknowledgements}}\nobreak
This work was supported in part by DOE grant DOE-91ER40618 and
by NSF PYI grant PHY-9157463.  I wish to thank T.~Banks, C.~Callan,
G.~Horowitz, J.~Preskill,
L.~Susskind L.~Thorlacius, and S.~Trivedi
 for useful conversations, and A.~Strominger for useful
conversations and
for comments on an earlier draft.

\listrefs

\end